# FINGER BASED TECHNIQUES FOR NONVISUAL TOUCHSCREEN TEXT ENTRY

MOHAMMED FAKRUDEEN, SUFIAN YOUSEF

*Department of Engineering & Built Environment, Faculty of Science and Technology, Anglia Ruskin University*
*Chelmsford, UK,*
*Email: mohammed.abdul@student.anglia.ac.uk, sufian.yousef@anglia.ac.uk*

MAHDI H.MIRAZ, ABDELRAHMAN HAMZA HUSSEIN

*College of Computer Sciences and Software Engineering,*
*University of Hail, KSA*
*Email :{m.miraz, ar.hussein}@uoh.edu.sa*

This research proposes Finger Based Technique (FBT) for non-visual touch screen device interaction designed for blind users. Based on the proposed technique, the blind user can access virtual keys based on finger holding positions. Three different models have been proposed. They are Single Digit Finger-Digit Input (FDI), Double Digit FDI for digital text entry, and Finger-Text Input (FTI) for normal text entry. All the proposed models were implemented with voice feedback while enabling touch as the input gesture. The models were evaluated with 7 blind participants with Samsung Galaxy S2 apparatus. The results show that Single Digit FDI is substantially faster and more accurate than Double Digit FDI and iPhone voice-over. FTI also looks promising for text entry. Our study also reveals 11 accessible regions to place widgets for quick access by blind users in flat touch screen based smartphones. Identification of these accessible regions will promote dynamic interactions for blind users and serve as a usability design framework for touch screen applications.

## 1. Introduction

With the help of assistive technologies such as screen readers, a blind user is able to access or input through a keyboard. As touch screens become increasingly dominant everywhere, accessing information through touch screen is becoming a major challenge for blind users. Blind users serving organizations such as the National Federation of the Blind [3], the American Foundation for the Blind [1], and the Royal National Institute of Blind People [2] suggest that blind people use iPhones as the touch screen device of choice.

   Text entry remains a time-consuming and error-prone accessibility challenge of touch screen devices. Studies on text entry on iPhone using the voice-over technique by Bonner *et al.* [7] and Oliveira *et al.* [4,9] found that mean entry speeds varied from 0.66 words per minute (WPM) to 2.1 WPM respectively. Other related works in this area emphasize the need for improved text entry in touch screen device for blind users.

The voice-over technique involves the hit and tries process by selecting the desired virtual keys for test entry. However, this paper focuses on an innovative technique based on hand position called Finger Based Technique (FBT) based on the work by [11]. In this technique, the locations of virtual keys on the touch screen are identified by the position of fingers on the touch screen by a blind user. The user sets the reference points at the tips of each finger.

Using FBT, *Finger-Digit Input*, a non-visual text entry method for digit was developed. In addition, *Finger-Text Input*, another non-visual text entry method for normal text entry, is also developed. In both methods, one hand is used to hold the touch screen and by using the other hand text entry is performed based on touch gesture.

Kane *et al.* [13] suggests that blind users do not need to know about spatial understanding while interacting with the touch screen. However, without spatial understanding, blind users took a lot of time and efforts to perform multi-touch gestures to accomplish tasks. These attempts cause fatigue and, overall, diminish the quality of work of blind users. Until now, an efficient way of text entry using a touch screen for blind users does not exist in the literature.

The main focus of this paper is to evaluate two proposed techniques: 1) to identify a technique for easy text entry by a blind user on the touch screen; and 2) to investigate obtained entry speeds using the identified technique and compare them with entry speeds using voice-over techniques. The result of our research will help to understand the spatial ability of blind users and lead to improvements in messaging systems.

## 2. Related Work

Literature on Human-Computer Interaction (HCI) proposes some non-visual touch screen text entry methods based on Braille. Oliver *et al.* [4,9] propose a new text entry based on Braille. The screen is divided into six virtual keys representing dots as in braille. The user types a character based on the specified region represented as dots. Braille type was found to be slower, having an entry rate of 1.45 WPM compared to Voice Over technique.

Frey *et al.* [5] suggest a text entry method called BrailleTouch. This technique is analogous to Braille type. However, in BrailleTouch the user has to input all the dots simultaneously to type a character. However, evaluation was not performed to measure user performance with BrailleTouch.

Mascetti *et al.* [6] revealed another braille based text entry method called TypeIn Braille. According to this technique, three simple gestures have to be performed to type a character. Measures assessing user performance with TypeIn Braille have not been conducted.

Non-Braille based techniques for non-visual text entry are also proposed in the literature. Bonner *et al.* [7] propose a non-visual text entry method named No-Look Notes where two virtual keys must be selected and activated to enter a character. The comparative evaluation of No-Look Notes with Voice-over found that the entry speeds were 0.66 WPM and 1.32 WPM respectively. Likewise, Sanchez and Aguayo [8] propose non-visual text entry method with virtual keys, however, without any assessment.

NavTouch is another non-visual text entry method proposed by Oliver *et al.* [9]. According to this technique, the user has to perform a sequence of gesture to choose (or to enter) a character. One evaluation revealed that the error rate was more than 10% and the typing speed was 1.72 WPM, which is much slower than the voice-over technique.

Tinwala and Mackenzie [10] and Yfantidis and Evreinov [12] have also proposed text entry methods through gestures. However, the methods were evaluated with blindfolded sighted users. Since the input ability of a sighted person varies significantly with blind people, their evaluation is not considered.

As mentioned earlier, the iPhone is found to be the most common touch screen device used by blind users and recommended by organizations for the blind. On iPhone, voice-over technique is used to access information by the blind user. When voice-over is running, a user can touch a virtual key to hear its audio feedback about the digit selected and double tap to activate that digit or a letter. As a result, voice-over enables eyes-free entry on the virtual QWERTY keyboard on the iPhone touch screen.

One of the important observations of this study is that it is imperative for developers to identify the accessible regions of touch screens for blind users. This will facilitate the performance of dynamic interactions such as those performed by sighted users. The existing related technologies for blind navigation lack this phenomenon. Without understanding the accessible regions and their access speed for flat touch screen surfaces, it will be difficult for developers to place the widget for accessing and to build an effective system in the future. This is, in fact, one of the key motivations for conducting this research study.

## 3. Rationale

One of the goals of our approach is to reduce the time for text entry in the touch screen. As mentioned earlier, a lot of time is consumed reaching the target character. With our techniques, many characters are aggregated into individual widgets. The widgets are placed near the position of the finger. This allows the user to set reference points near to the position held by the hand. As a result, this technique enhances speedy interaction by blind users with the touch screen. Additionally, the goal of designing FBT prototypes for text entry is to improve usability in terms of users' ability, preferences, and variation in screen sizes (e.g. smartphones to tablets).

Using current technologies such as the voice-over technique, a blind user uses audio feedback when touching to explore required virtual keys, and then activates the key. These processes increase the duration of text entry. Alternatively, large virtual keys were used for quick exploration [8]. However, only a few keys fit into a small screen. Consequently, FBT technique is intended to execute the text entry task faster by having the users interact using predefined identifiable regions.

Finally, the chording nature and algorithm were used to make text entry easier. The complexity involved in multi-touch gestures makes input difficult for blind users. However, the FBT technique implements only touch gestures as the means of input for text entry. This approach alerts users to correct text input errors and to read what has been typed.

Current technologies implement a static layout for blind users to interact with touch screen devices. This results in blind users recognizing two-dimensional pages as only being a single horizontal list that includes a large set of items. Remembering the larger sequence of items is an additional burden.

On the whole, dynamic interactions are preferred by sighted users to recognize items using vision. In contrast, blind users use gestures such as touch or flicker to recognize items in a static layout. As a consequence, the spatial understanding about the arrangement of items is lost. Furthermore, distinct navigation has to be maintained separately for blind users. Comparatively, the requirement and usability issues vary for blind and sighted users for the same structure of a single page layout. There is a need to improve accessibility and usability in the area of text entry in order to allow blind users performance to be similar to or approach that of sighted users.

In order to achieve the goal of parity between blind and sighted users, this chapter will extend the previous study by [11] on FBT. It also focuses on extending FBT to design, develop and evaluate a prototype for text entry that is authentic and timely.

## 4. Finger Based Entry Method

Text entry on touch screen smartphones by blind users can be made more proficient based on the following factors:

1) Spatial representation of virtual keys
2) Audio feedback upon tapping/touching the buttons
3) Number of virtual keys and
4) The mechanism provided to edit any mistakes committed.

This paper mainly focuses on the design and development of two techniques. Both techniques are designed based on FBT. In the first prototype, the text entry was digit based on phone dialler application. The second prototype was based on normal text entry.

### 4.1. *Finger-Digit Input (FDI) Entry Method*

The FDI method consists of two models: Single Digit FDI (as shown in Fig. 1a) and Double Digit FDI (as shown in Fig. 1b). A virtual key relates to a Single Digit in the Single Digit FDI. Alternatively, in Double Digit FDI, each virtual key links to two different digits. Each model provides voice feedback on touch/tap event performed by the blind users.

In both cases, as shown in Fig. 1a and Fig. 1b, the user holds the device in the left hand. One side of the device is held by the index finger, the middle finger, the ring finger and the little finger of left hand. The left hand thumb finger holds on to the other side of the device. Using right hand fingers, the user taps the buttons.

The blind user assumes that the regions adjacent to the tips of left hand fingers represent the associated virtual digit. For instance, in the Single Digit FDI technique, as shown in Fig. 1a, if the user presses the region near to the middle finger, the virtual key representing the $2^{nd}$ digit is invoked. Table 1 shows the list of virtual digital keys along with their respective finger holding positions.

Table 1. Position of widgets in FDI and FTI prototype.

| POSITION OF BUTTONS WITH RESPECT TO HOLDING LEFT HAND FINGER | SINGLE DIGIT FDI | DOUBLE DIGIT FDI | FTI |
| --- | --- | --- | --- |
| Above index finger | Backspace | Backspace | Delete |
| Index finger | Four | One and Two | ABCD |
| Middle finger | Five | Three and | EFGH |
| Ring finger | Six | Four Five and Six | IJKL |
| Little finger | Seven | Seven and Eight | 123 |
| Below Little finger | Eight | Nine and Ten | UCASE/ LCASE |
| Center of touch screen | Nine | | ENTER |
| Thumb | Two | Enter | QRST |
| Above the thumb | One | | MNOP |
| Below the thumb | Three | Contacts | UVWXYZ |
| Bottom Center of the touch screen | Call | Call | SPECIAL CHARACTERS |
| | Backspace | Backspace | SEND |

In Double Digit FDI, the blind user accesses the proper region by holding the tip of each finger in the area that is according to the two digits each finger represents. For example, the index finger represents the digits "1" and "2", so the tip of the index finger will be held in the region the "12" button (Fig. 1b). A single press will temporarily add the first digit to dialling number textbox. For adding the second number instead, two subsequent presses are required .

For instance, to add "2" to the dialling number text box, a blind user will press the region near to the tip of index finger to invoke the "12" button. Upon the first press, the audio informs the users that "1" has been pressed. Upon the second press, the audio informs the users that "2" has been pressed. Now the users have to press the "Enter Key" to finally add "2" to the dialling number textbox. If the users want to add "1" instead, they have to press the "Enter Key" immediately after the pressing the "12" button for the first time.

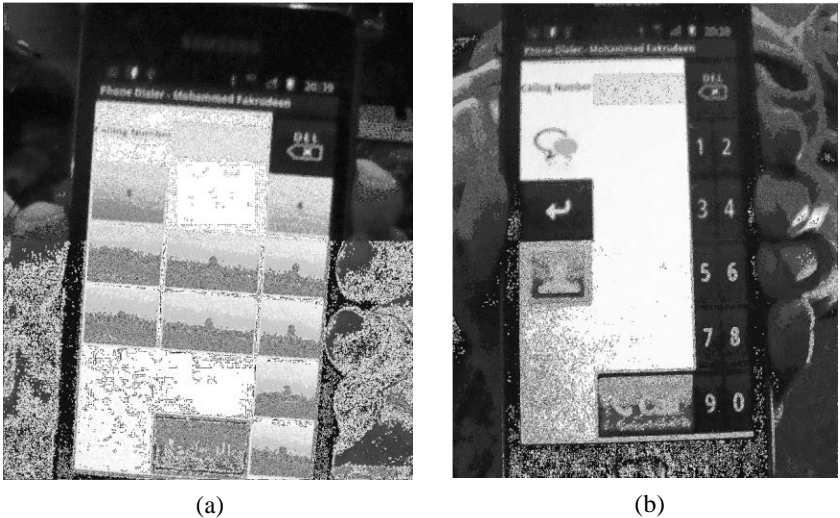

(a) (b)
Figure 1. (a) Single Digit FDI (b) Double Digit FDI.

### 4.2. *Finger-Text Input Text Entry Method*

Finger-Text Input (FTI) uses Finger Based Techniques as proposed by Fakrudeen et al. [11]. This technique was evaluated for text entry based on finger position of a blind user over the smartphone. The blind user assumes the character "ABCD" are near to the index finger, "EFGH" to the middle finger, "IJKL" to the ring finger, numbers from 1-10 to the little finger. In the same manner, "QRST" was assumed to be near to the thumb, "MNOP" above the thumb and, "UVWXYZ" below the thumb(see Fig. 2). The extra characters will be located at the middle of the index and thumb fingers. Below the little finger, there will be a mode to change from upper to lower and vice versa. The virtual key above the center button of a smartphone will be the Enter key to enter characters in the text box. For instance, if a user wants to type "S", the user presses the button near to Finger 1, 3 times and then presses the Enter key to add the character to the text box.

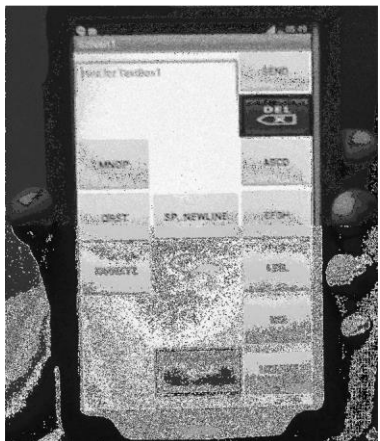

Figure 2. FTI prototype.

## 5. Study 1: Finger-Digit Input Entry Method

The user evaluation was carried out to evaluate the Finger-Digit Input and Finger-Text Input techniques for touch screen based smartphones with blind users.

### 5.1. *Method*

#### 5.1.1. *Participants*

With an average age of 35 years, we recruited seven blind participants. None of the participants had previous experience of using a touch screen mobile. However, they had enough experience of using mobiles with screen readers. Due to small cohort size, we adopted the "repeated trial" methodology and conducted 8 trials per participant.

#### 5.1.2. *Apparatus*

The prototypes were developed using Android 4.0 Ice Cream Sandwich. Both the prototypes can run on any mobile phone that supports the Android platform. Prototypes were tested on the Samsung Galaxy S2. No additional hardware is required to run them. During the development phase, the prototypes were tested on an emulator supplied by Android SDK. After successful completion of the development phase, they were deployed on the Samsung Galaxy S2 mobile phone and tested with the blind users.

5.1.3. *Procedure*

The study includes two sessions: training and evaluation. During the training session, the participants were shown how to use the prototypes explaining the difference between Single Digit FDT and Double Digit FDT.

The participants were then asked to perform small tasks such as dialing a number. After successfully dialing the number, a second number was provided for the blind users to dial in order to increase proficiency and ease prior to evaluation.

During the evaluation session, each participant was given eight different contact numbers. Each number consisted of 10 digits. For the comfort of blind participants, one number was provided at a time.

**5.2. *Results***

The experiment was analyzed based on: 1) Prescribed text (P) - the text to be typed by participant, 2) Transcribed text (T) - the text typed by participants and 3) duration to type each phrase. This also involved the number of wrong digits entered and the number of corrections performed by the participants. For both single and Double Digit FDI, the experiment was tested for the following factors:

1. Entry Speed – Words per Minute
2. Performance over duration
3. Error Rates and
4. Preferences

The last factor was evaluated using a Likert Scale [1-Strongly Disagree, 5-Strongly Agree] based on the responses collected through questionnaires and data related to other factors that were recorded and collected during the evaluation process.

5.2.1. *Words per Minute (WPM)*

WPM measures the time taken to produce target digit. As the aim of our research is not to build the messaging system, but to adopt the process of text entry, we assume digit entry and text entry to be similar process. Thus, the WPM text entry metric is adopted for our technique. WPM does not consider gesture. It considers the length of target number and the number of keystrokes made during digit entry. WPM is defined as [12]:

$$WPM = [(|T| - 1) / S] * 60 * (1/5)$$

Here, S is the time taken to, enter the target number in seconds and |T| is the length of targeted number.

The examination of entry speed assesses the ability of blind users in accessing the location point according to finger positions.

WPM is checked for normality by using the Shapiro Wilkson (W) test technique. The data are normalized for Single Digit FDI using WPM (W (6) =0. 972, P<0.05) and Double Digit FDI (W (6) =0. 896, P<0.05). Hence, one-way ANOVA was performed for normalized data to find the significance of each technique on the WPM.

Based on the result obtained it can be concluded that there is a statistically significant difference between Single Digit and Double Digit FDI as determined by one-way ANOVA ($F$ (1,10) = 6.600, $p =$. 028). It can be further concluded that the text entry speed (WPM) of Single Digit FDI (Mean=3. 26, SD=0. 784) was faster than that of the Double Digit technique (Mean =1. 92, SD=1. 00) as shown in Fig. 3.

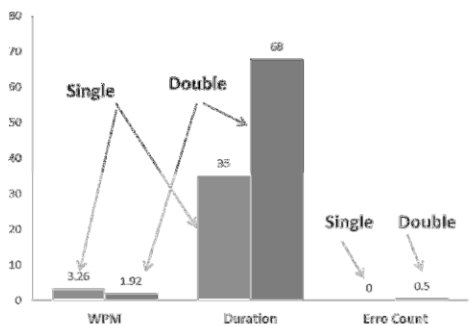

Figure 3. Mean of dependent variables.

The achieved entry speed of the Double Digit FDI is slower as compared with Single Digit FDI is attributed to the following:
1. Participants have to press the button two times for digits such as 2,4,6,8,0 and
2. Participants have to press the enter key for confirmation (to add an entry to the textbox).

It can be concluded that the Single Digit FDI technique is more promising than Double Digit FDI for the purpose of digit input.

### 5.2.2. *Performance over Duration*

The duration can be defined as the time taken to complete dialling a phone number. The phone numbers provided consisted of 10 digits each. The entry speed was examined over the entire duration of the test.

The duration is checked for normality by using the Shapiro Wilkson (W) test technique. The duration is normalized for Single Digit FDI using duration (W (6) =0. 972, P<0.05) and Double Digit FDI uses duration (W (6) =0. 896, P<0.05). Therefore, one-way ANOVA was performed to find the significance of each technique on the duration.

There is a statistically significant difference between Single Digit and Double Digit FDI as determined by one-way ANOVA ($F$ (1,10) = 4.499, $p$ <0.05). As a result, we can conclude that the time taken for text entry using Single Digit FDI (Mean=35.00, SD=9.67) was less than the time taken for the Double Digit technique (Mean =68.50, SD=37.45).

### 5.2.3. *Error Count*

The number of digit(s) entered incorrectly was also analyzed. Out of seven participants, only two of them had to perform digit correction while using Double Digit FDI (Fig. 4).

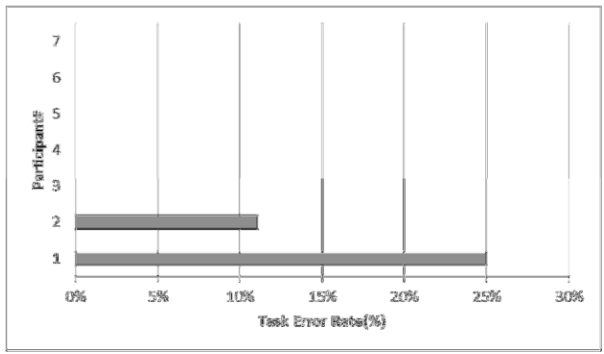

Figure 4. Task Error Rate for Double FBT.

The error count is checked for normality by using the Shapiro Wilkson (W) test technique. The number of errors is not normalized for Double Digit FDI (W Doub(6) =0. 701, P=0. 006). Hence the Mann-Whitney U test was performed for non-normalized data to find the significance of each technique on the error count.

There was no statistically significant difference between the error count of the techniques (U = 12, P =0. 14). As a result, a firm conclusion cannot be made. This is an area that will demand further research in the future.

### 5.2.4. *Preference*

Although the entry speed of Single Digit FDI was faster than Double Digit FDI, 70% of the participants prefer Double Digit FDI as compared to Single Digit FDI (as shown in Fig. 5a).

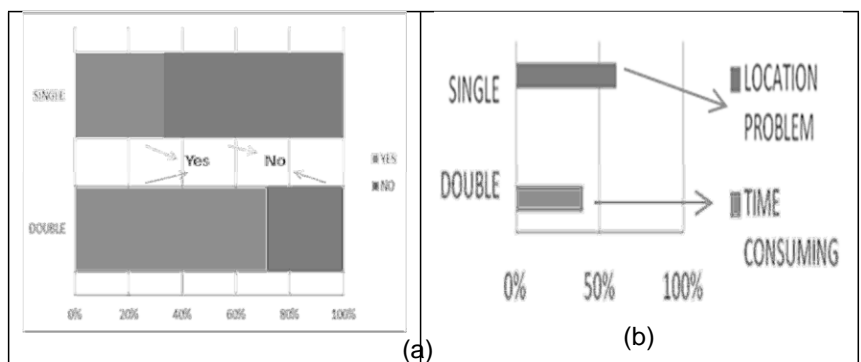

Figure 5. (a) Desire's of FDI. (b) Problems related to FDI.

This was attributed to the following factors (as shown in Fig. 5b):
1. *Location problem*: Only 5 buttons are placed in 'Double Digit' FDI for digits as compared to 10 buttons for 'Single Digit' FDI. 60% of the participants found it to be difficult to locate the buttons in 'Single Digit' FPD.
2. *Time Consuming*: Only 40% of blind users are of the opinion that the 'Double Digit' FDI is time consuming as compared to 'Single Digit' FDI because:
   a. They had to press the same button twice for certain digit such as 2,4,6,8 and 0.
   b. They also had to press the enter key for confirmation of adding each digit.

## 6. Study 2: Finger-Text Input Entry Method

### 6.1. *Procedure*

This study included two sessions: training and evaluation. During the training session, the participants were shown how to use the prototypes explaining the difference between FDI and FTI.

The participants were then asked to perform small tasks such as typing a word. After being able to type a word, another word was provided to the blind user to achieve proficiency and ease.

During the evaluation session, each participant was given eight different phrases. Each phrase consisted of 9-20 characters. For the comfort of blind participants, one phrase was provided at a time.

### 6.2. *Results*

The results reveal that mean entry speed for FTI was 7.02 WPM, which is higher than [4,9]. The average time taken for text entry is 70.02. Since FTI is the first application the text entry, it can be improved in many ways. We plan to compare the FTI text entry with voice-over. In addition, we also plan to detect Maximum Likelihood location for the characters in the available region/slot to improve WPM measure. Thus, this research reveals that text entry with FTI is promising and it has to be improved for more efficiency.

## 7. Discussion

The performance of blind participants at the initial stage was not as profound as expected. The users' level of inexperience with touch screen technology was a major challenge during the study. However, user performance did demonstrate noticeable improvement during the training sessions. Conducting training sessions is resultantly proven to be decidedly advantageous.

The average entry speed of Single Digit FDI is 3.26 WPM, which is much higher than the mean entry speed on iPhone voice-over(2.1 WPM) by Oliver *et al.* WPM. On the contrary, the mean entry speed of Double Digit FDI is 1.93, which is lower than Oliver's voice-over entry speed. The average entry speed is much higher than the voice-over entry speeds in related work [4,7, and 9]. The average speed of FTI is also much higher than the findings in [4, 7, and 9]. As a result, Single Digit FDI outperforms voice-over in digital entry, the *de facto* eyes-free digital entry method for touch screen devices. The FTI also shows promise for text entry by blind users.

Our research provides insight into the performance of our technique by blind users using touch screen devices. The research reveals that blind users can identify 11 reference regions for quicker accessibility and navigation for flat touch screen devices [11]. For better interaction, identified accessible regions will be more appropriate to place the widget. In addition, it guides the developer to understand the design pattern for touch screen applications used by blind users.

Both FDI and FTI techniques depend on the overlays at the edge of the touch screen where the blind user holds the device. We tested both of our applications using different devices such as tablets and smartphones. Interestingly, all the buttons lay on the edge at the finger holding position. Thus, our application holds promise for cross-platform compatibility. However, it needs more analysis and testing.

Our findings are specific to smartphones and other mobile devices. These findings establish the reliability of our techniques for accessing the flat touch screen without any additional assistive device.

## 8. Conclusion

Our work suggests an innovative interaction technique for flat touch screens to be used by the blind users. The proposed technique encourages developers to place the widget properly to enable blind users to experience dynamic interaction with touch screen devices. The evaluation shows that our technique is much faster than iPhone voice-over entry speeds in performing similar tasks.

The paper also focused on identifying the accessible regions for dynamic interaction. By properly adopting these techniques, non-functional requirements such as system usability can be achieved.

Since FDI and FTI are the first application of the Finger-Based Technique, it can be enhanced in many ways in the future. The FTI can also be remodeled for Braille users. Using current techniques, text entry by blind and sighted user differs widely. Thus, our research narrows the gap between sighted and blind users for text entry methods.

## References

1. D. Burton, "Cell Phone Access: The Current State of Cell Phone Accessibility", [Online]. Available: http://www.afb.org/afbpress/pub.asp?DocID=aw120602 (2011).
2. RNIB, "A guide to talking mobile phones and mobile phone software," [Online]. Available:

http://www.rnib.org.uk/livingwithsightloss/Documents/Mobile_phone_software_factsheet_PDF.pdf. [Accessed 02 03 2014].
3. NFB, "How many children in America are not taught to read?", [Online]. Available: https://nfb.org/braille-initiative (2013) [Accessed 05 03 2014].
4. J. Oliveira, T. Guerreiro, H. Nicolau, J. Jorge and D. Gonçalves, "BrailleType: unleashing braille over touch screen mobile phones," in *INTERACT'11 Proceedings of the 13th IFIP TC 13 international conference on Human-computer interaction - Volume Part I* (2011).
5. B. Frey, C. Southern and M. Romero, "Brailletouch: mobile texting for the visually impaired," in *UAHCI'11 Proceedings of the 6th international conference on Universal access in human-computer interaction: context diversity - Volume Part III* (2011).
6. S. Mascetti, C. Bernareggi and M. Belotti, "TypeInBraille: a braille-based typing application for touchscreen devices," in *ASSETS '11 The proceedings of the 13th international ACM SIGACCESS conference on Computers and accessibility*, Dundee, Scotland (2011).
7. M. N. Bonner, J. T. Brudvik, G. D. Abowd and W. K. Edwards, "No-look notes: accessible eyes-free multi-touch text entry," in *Proceeding Pervasive'10 Proceedings of the 8th international conference on Pervasive Computing* (2010).
8. J. Sánchez and F. Aguayo, "Mobile messenger for the blind," in *ERCIM'06 Proceedings of the 9th conference on User interfaces for all*, Bonn, Germany (2006).
9. J. Oliveira, T. Guerreiro, H. Nicolau, J. Jorge and D. Gonçalves, "Blind people and mobile touch-based text-entry: acknowledging the need for different flavors," in *ASSETS '11 The proceedings of the 13th international ACM SIGACCESS conference on Computers and accessibility*, Dundee, Scotland (2011).
10. H. Tinwala and S. S. Mackenzie, "Eyes-free text entry on a touchscreen phone," in *IEEE Toronto International Conference Science and Technology for Humanity (TIC-STH)*, Toronto (2009).
11. M. Fakrudeen, S. Yousef and M. H. Miraz, "Finger Based Technique (FBT): An Innovative System for Improved Usability for the Blind Users' Dynamic Interaction with Mobile Touch Screen Devices," *Lecture Notes in Engineering and Computer Science: Proceedings of The World Congress on Engineering 2014, WCE 2014, 2-4 July, 2014, London, U.K.,* pp. 128 – 133(2014).
12. G. Yfantidis and G. Evreinov, "Adaptive blind interaction technique for touchscreens," *Universal Access in the Information Society,* vol. 4, no. 4, pp. 328 – 337 (2006).
13. S. K. Kane, J. P. Bigham and J. O. Wobbrock, "Slide rule: making mobile touch screens accessible to blind people using multi-touch interaction techniques," Halifax, Nova Scotia, Canada (2008).